\documentclass[journal]{IEEEtran}

\usepackage{cite}
\usepackage{hyperref}

\usepackage{multicol}

\ifCLASSINFOpdf
   \usepackage[pdftex]{graphicx}
\else
   \usepackage[dvips]{graphicx}
\fi
\usepackage{amssymb}

\usepackage[cmex10]{amsmath}
\usepackage{algorithm}
\usepackage{multirow}

\usepackage{algorithmic}
\usepackage[tight,footnotesize]{subfigure}
\usepackage{url}

\begin{document}
%
\title{Full Duplex Wireless Communications for Cognitive Radio Networks}

\author{\IEEEauthorblockN{Wenchi Cheng$^{1}$, Xi Zhang$^{1}$, and Hailin Zhang$^{2}$}~\\[0.2cm]

\IEEEauthorblockA{$^{1}$Networking and Information Systems Laboratory\\
Dept. of Electrical and Computer Engineering, Texas A\&M University, College Station, TX 77843, USA\\
$^{2}$State Key Laboratory of Integrated
Services Networks, Xidian University, Xi'an, China\\
E-mail: \{\emph{wccheng@tamu.edu}, \emph{xizhang@ece.tamu.edu}, \emph{hlzhang@xidian.edu.cn}\}}

\thanks{This work is supported by the U.S. National Science Foundation CAREER
Award under Grant ECS-0348694, the 111 Project in Xidian University of China (B08038),
National Natural Science Foundation of China (No.61072069),
the Fundamental Research Funds for the Central Universities (72101855\&72105242),
the Natural Science Basic Research Plan in Shaanxi Province of China (No.2010JM8001).}

\thanks{The major contents of this technical report will appear in IEEE Information Society CISS 2011.}
}


%


\maketitle

\begin{abstract}
As a key in cognitive radio networks (CRNs), dynamic spectrum access needs to be carefully designed to minimize the interference and delay to the \emph{primary}~(licensed) users. One of the main challenges in dynamic spectrum access is to determine when the \emph{secondary}~(unlicensed) users can use the spectrum. In particular, when the secondary user is using the spectrum, if the primary user becomes active to use the spectrum, it is usually hard for the secondary user to detect the primary user instantaneously, thus causing unexpected interference and delay to primary users. The secondary user cannot detect the presence of primary users instantaneously because the secondary user is unable to detect the spectrum at the same time while it is transmitting. To solve this problem, we propose the full duplex wireless communications scheme for CRNs. In particular, we employ the Antennas Cancellation (AC), the RF Interference Cancellation (RIC), and the Digital Interference Cancellation (DIC) techniques for secondary users so that the secondary user can scan for active primary users while it is transmitting. Once detecting the presence of primary users, the secondary user will release the spectrum instantaneously to avoid the interference and delay to primary users. We analyze the packet loss rate of primary users in wireless full duplex CRNs, and compare them with the packet loss rate of primary users in wireless half duplex CRNs. Our analyses and simulations show that using our developped wireless full duplex CRNs, the packet loss rate of primary users can be significantly decreased as compared with that of primary users by using the half duplex CRNs.
\end{abstract}

\begin{IEEEkeywords}
Cognitive radio networks (CRNs), full duplex CRNs, dynamic spectrum sensing, packet loss rate, interference control.
\end{IEEEkeywords}


\section{Introduction}
\IEEEPARstart{C}{ognitive} radio has been an effective method to solve the problem of low spectrum utilization, which is caused by the current fixed frequency allocation policies~\cite{FCC_2003}. In the cognitive radio networks~(CRNs)\cite{Hang_CISS_2007}, a challenging problem is that the secondary users~(SUs) need to scan and identify the spectrum state to verify whether the spectrum is used by the primary users~(PUs) or not. In particular, when the SU is transmitting, it is usually hard for the SUs to identify the presence of the PUs immediately, thus causing unexpected interference and delay to PUs. This is because the SUs work in the wireless half duplex fashion, and thus cannot transmit and receive signals simultaneously.

Under the constraint of half duplex, previous works mainly focused on using a periodic sensing scheme for SUs to periodically scan whether the PUs are presence again or not~\cite{Cross_JSAC_XiZhang_2008,Peng_2007_WCNC,Liang_2008_TWC}. Obviously, a too long detection time will deteriorate the throughput of the SUs while a too short detection time will increase the missed detection probability of SUs' sensing for the active PUs. Therefore, we believe that the thorough solution for the SUs to timely and efficiently detect the PUs is for SUs to transmit while listening to the channel simultaneously, which implies that the SUs need to work in the wireless full duplex fashion.

The reason why the full duplex cannot be used in wireless communications is that there is a large power difference between the local transmitted signals and the received signals from the other nodes, which makes it hard to subtract the local transmitted signals from the received signals. Recently, some research works have shown the possibility of using full duplex in wireless communications~\cite{Quellan_Inc,Radunovic_WIMESH_2010,Gollakota_Sigcomm_2008,Choi_2010_Mobicom}. By combining the Antenna Cancellation (AC), the RF Interference Cancellation (RIC), and the Digital Interference Cancellation (DIC) techniques, the full duplex can be used in wireless communications~\cite{Choi_2010_Mobicom}. The full duplex transmission mode can also have a wide range of networking applications, such as mobile multicast networks~\cite{XiZhang_Multicast_2002,XiZhang_Multicast_2004,XiZhang_Flow_2003}.

In this paper, we propose and evaluate the wireless full duplex scheme for CRNs. We characterize the PUs' packet loss rate in wireless full duplex CRNs and wireless half duplex CRNs. Considering that a number of factors can lead to the imperfectness of full duplex, we also analyze the effect of imperfect full duplex wireless communications in CRNs. We show that by using full duplex in CRNs, the packet loss rate of PUs can be decreased significantly as compared with  that in half duplex CRNs.

The rest of this paper is organized as follows. Section~\ref{System Model} describes the architecture of wireless full duplex nodes and the CRN model. Section~\ref{Performance Analysis of Wireless Full and Half Duplex CRNs} analyzes and compares the PUs' packet loss rate in the wireless half duplex CRNs, the perfect wireless full duplex CRNs, and the imperfect wireless full duplex CRNs. Section~\ref{Simulation Results} evaluates our proposed wireless full duplex in CRNs. The paper concludes with Section~\ref{Conclusions}.

\section{System Model}
\label{System Model}
\subsection{The Wireless Full Duplex Node Architecture}
In wireless communications, because of the serious fading over wireless channels, the signal from a local transmit antenna is hundreds of thousands of times stronger than the signal received from the other nodes. Hence, it has been generally assumed that the wireless CRN node cannot decode a received signal at a radio at the same time while it is simultaneously transmitting.

\begin{center}
\begin{figure}[!htb]
\centering
\includegraphics[width=2.8 in]{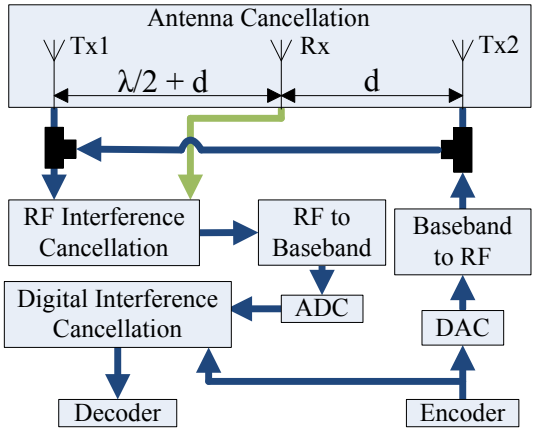}
\caption{The Wireless Full Duplex Node Architecture.}
\label{Full_Duplex_Node}
\end{figure}
\end{center}

However, in principle it is possible to build the full duplex in wireless systems. Since the system knows the transmit antenna's signal, it can subtract the transmit signal from the receive antenna's signal and decode the remainder using standard techniques. The main factor preventing the full duplex in wireless system from its implementation is the large power difference between the local transmit signal and receive signal which is sent by the other nodes. Therefore, once we remove the large power difference by some reduction measures, we can realize the full duplex transmission in wireless systems. Some works have revealed the possibility of using full duplex transmission in wireless systems
\cite{Quellan_Inc,Radunovic_WIMESH_2010,Gollakota_Sigcomm_2008,Choi_2010_Mobicom}. In this paper, we propose the scheme of combining using the Antenna Cancellation, the RF Interference Cancellation, the Digital Interference Cancellation techniques in CRNs. Fig.~\ref{Full_Duplex_Node} shows the block of diagram of a SU incorporating all the three techniques for full duplex operation. In particular, for a wavelength $\lambda$, two transmit antennas are placed at $d$ and $\lambda/2+d$ away from the receive antenna, respectively. Offsetting the two transmitters by half a wavelength causes their signals to add destructively. This creates a \emph{null position} where the receive antenna hears a much weaker signal (self-interference) compared with any one of the local transmit signals. After the Antenna Cancellation, the self-interference has been attenuated to a low enough level. Then, we can use the RF Interference Cancellation and the Digital Interference Cancellation techniques to further decrease the self-interference.

\subsection {Cognitive Radio Network Model}
\begin{center}
\begin{figure}[!htb]
\centering
\includegraphics[width=3.4 in]{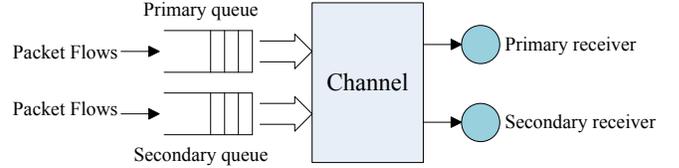}
\caption{Single Channel Cognitive Radio Network Model.}
\label{Single_Channel_CRNs}
\end{figure}
\end{center}

In this paper, we consider a cognitive radio network with two input flows as illustrated in Fig.~\ref{Single_Channel_CRNs}. Both of the arrived rate of PUs' and SUs', denoted by $\lambda_\mathrm{p}$ and $\lambda_\mathrm{s}$, respectively, are assumed to be independent Poisson processes. The PUs' flow has \emph{preemptive} priority over the SUs' flow. If one packet arrives into the system and cannot be transmitted immediately, it will be stored in the corresponding buffer in a First-In-First-Out (FIFO) manner where the buffer is assumed to be large enough and therefore no packet will be dropped due to the overflow.

\section{Analyses of Wireless Full and Half Duplex CRNs}
\label{Performance Analysis of Wireless Full and Half Duplex CRNs}
Minimizing the interference caused by SUs to PUs is a challenging problem in CRNs. When the PU becomes active to use the channel occupied by the SU, to minimize the SUs' interference to PUs, the SU must release the channel immediately, which implies that the SU can constantly sense the channel when it is transmitting on the same spectrum. However, as wireless signals attenuate quickly over distance, the signal from the SU's local transmit antenna is hundreds of thousands of times stronger than the signal received from the PU. Therefore, in general SUs cannot decode a received signal while they are transmitting simultaneously.

Since the node knows the local transmit signal, it can subtract the local transmit signal from the receive antenna's signal. The main problem is how to cancel the local transmit signal (self-interference) signal. We propose to employ the AC, the RIC, and the DIC techniques in CRNs \cite{Choi_2010_Mobicom}. These techniques enable the SUs to work in the wireless full duplex fashion.

Assuming every PU and SU has $N$ packets to transmit, the packet length is denoted by $L_{p}$, and the channel data rate is denoted by $R$. We also assume that once the PU wants to access the channel again, it can tolerate a delay for a period less than $D$. If the delay is larger than $D$, the packets of PU will be dropped due to quality of service (QoS)~\cite{Jia_TWC_2007,Jia_TWC_2_2007,XiZhang_Mag_2006,JiaTang_TWC_2007,JiaTang_TWC_2008,JiaTang_JSAC_2007} requirement.

In ideal wireless full duplex CRNs, we do not consider the missed detection (SUs fail to detect the activation of PU) and the imperfect of full duplex. This implies that the packet loss rate of PUs is zero. We denote $N_{SU}$ as the number of packets successfully transmitted by the SUs during the time denoted by $T_{w}$. Therefore, during the time $T_{w}$, the packet loss rate of PUs in half duplex CRNs, denoted by $L_{PU}$, can be written as follows:
\begin{equation}
\label{Half-Duplex-Frame-Loss-NO}
\begin{aligned}
L_{PU} = \frac{N_{SU}\bigg[\sum\limits_{q=1}^{\infty}q\frac{(\lambda{p}\delta_{t})^{q}}{q!}e^{-\lambda{p}\delta_{t}}\bigg]
}{N\lambda_{p}T_{w}}
= \frac{N_{SU}\delta_{t}}{NT_{w}},
\end{aligned}
\end{equation}
where $\delta_{t}$ is equal to $(NL_{p}/R)-D$. Eq.~\eqref{Half-Duplex-Frame-Loss-NO} implies that if any PU is active after a SU's transmission starting time, but not later than $\delta_{t}$ from the SU's transmission starting time, the packets of the PU will be dropped.

In practice, it is impossible for the SUs to sense the presence of the PUs without any error. Several signal detection techniques can be used for spectrum sensing, such as the energy detection, feature detection, and matched filter, for the SUs to detect the presence of the PUs \cite{Cabric_2004_Asilomar}. We focus on the energy detection approach in this paper because the energy detection approach is efficient and simple to be implemented in hardware, and more importantly, it does not require the knowledge of signal features of the PUs, which typically may not be known to the SUs.

Taking the imperfect spectrum sensing into consideration, the packet loss rate of PUs in the full duplex CRNs, denoted by $\rho_{PU}$, can be written as follows:
\begin{equation}
\label{Imperfect-Sensing-Full-Duplex-Frame-Loss-NO}
\begin{aligned}
\rho_{PU} = p_{MD},
\end{aligned}
\end{equation}
where $p_{MD}$ is the SUs' missed detection probability. The packet loss rate of PUs in half duplex CRNs, denoted by $h_{PU}$, can be written as follows:
\begin{eqnarray}
\label{Imperfect-Sensing-Half-Duplex-Frame-Loss-NO}
h_{PU} &= 1-(1-L_{PU})(1-p_{MD})\nonumber \\
&= \frac{N_{SU}\delta_{t}}{NT_{w}}(1-p_{MD})+p_{MD}.
\end{eqnarray}
We assume that the data of SUs can tolerate unlimited delay, thus $N_{SU}$ can be calculated by $N\lambda{s}T_{w}$. Now, the main problem is how to find the closed-form of $p_{MD}$. The missed detection probability in imperfect full duplex CRNs can be larger than the missed detection probability in half duplex CRNs. We first assume the full duplex is perfect, and then consider the case of imperfect full duplex.

\subsection{The Perfect Full Duplex CRN}
To obtain the closed-form expression of $p_{MD}$, we need to analyze the received signal at the receive antenna. After the AC, the received signal, denoted by $r(t)$, can be written as follows:
\begin{equation}
\label{Received-Signal}
\begin{aligned}
r(t)=&hs(t)+\omega(t)+\kappa(t),
\end{aligned}
\end{equation}
where $h$ is the instantaneous amplitude gain of the channel between the SU and the PU, which follows Rayleigh distribution, $s(t)$ is the PU's sent signal with transmit power $E_{s}$, $\omega(t)$ represents the additive white Gaussian noise (AWGN) with zero mean and variance of $\sigma^{2}$, and $\kappa(t)$ denotes the self-interference.

Then, by using the RIC and the DIC, we can obtain additional SNR gains about 10dB and 20dB, respectively. In this paper, we mainly focus on the impact of using wireless full duplex in CRNs. Hence, we mainly analyze the effect of using the AC in CRNs, while assuming the SNR gains of the RIC and the DIC are ideal values as 10dB and 20dB, respectively.

In the wireless full duplex CRNs, using the AC technique, we can derive the self-interference as follows:
\begin{eqnarray}
\label{Self-Inteference}
\kappa(t)&=&A_{ant}e^{j\left(2\pi f_{c}t+\phi_{1}\right)}+\left(A_{ant}
+\epsilon^{A}_{ant}\right)\nonumber\\
&&\cdot e^{j\left(2\pi f_{c}t+\phi_{1}+\pi+\frac{2\pi\epsilon^{d}_{ant}}{\lambda}\right)},
\end{eqnarray}
where $A_{ant}$ is the amplitude of signal at the receive antenna received from a single transmit antenna, $\phi_{1}$ is the signal phase shift from transmit antenna to receive antenna, $\epsilon^{A}_{ant}$ represents the amplitude difference between the received signals from the two transmit antennas at the receive antenna, $f_{c}$ denotes the carrier frequency, $\epsilon^{d}_{ant}$ is the phase error caused by receive antenna placement compared to the ideal case where the signals from the two antennas arrive with $\pi$ out of phase of each other, and $\lambda$ denotes the wavelength related to the carrier frequency.

Substituting Eq.~\eqref{Self-Inteference} into Eq.~\eqref{Received-Signal}, we get
\begin{eqnarray}
r(t)&=&hs(t)+\omega(t)+A_{ant}e^{j\left(2\pi f_{c}t+\phi_{1}\right)}\nonumber\\
&&+\left(A_{ant}+\epsilon^{A}_{ant}\right)
e^{j\left(2\pi f_{c}t+\phi_{1}+\pi+\frac{2\pi\epsilon^{d}_{ant}}{\lambda}\right)}.
\end{eqnarray}

Thus the objective of spectrum sensing is to decide between the following two hypotheses ($\mathcal{H}_{1}$, $\mathcal{H}_{0}$) as shown in Eq.~\eqref{Hypo}, where $\mathcal{H}_{1}$ is the hypothesis stating that the idle PU becomes active again, and $\mathcal{H}_{0}$ is the hypothesis stating that the PU is not active.
\begin{figure*}[!t]
\begin{equation}
\label{Hypo}
r(t)=\left\{
\begin{array}{ll}
hs(t)+\omega(t)+A_{ant}e^{j\left(2\pi f_{c}t+\phi_{1}\right)}
+\left(A_{ant}+\epsilon^{A}_{ant}\right)
e^{j\left(2\pi f_{c}t+\phi_{1}+\pi+\frac{2\pi\epsilon^{d}_{ant}}{\lambda}\right)}, &\mathrm{if}\ \mathcal{H}_{1} \\
\\
\omega(t)+A_{ant}e^{j\left(2\pi f_{c}t+\phi_{1}\right)}
+\left(A_{ant}+\epsilon^{A}_{ant}\right)
e^{j\left(2\pi f_{c}t+\phi_{1}+\pi+\frac{2\pi\epsilon^{d}_{ant}}{\lambda}\right)}, &\mathrm{if}\ \mathcal{H}_{0}
\end{array}
\right.
\end{equation}
\hrulefill
\vspace*{4pt}
\end{figure*}

In the case of perfect full duplex, since both $\epsilon^{A}_{ant}$ and $\epsilon^{d}_{ant}$ are zero, Eq.~\eqref{Self-Inteference} equals zero, causing no self-interference on received signal at receive antenna. Therefore, Eq.~\eqref{Hypo} can be reduced to:
\begin{eqnarray}
\label{Idea}
r(t)=\left\{
\begin{array}{ll}
hs(t)+\omega(t), &\mathrm{if}\ \mathcal{H}_{1}\\
\\
\omega(t), &\mathrm{if}\ \mathcal{H}_{0}
\end{array}
\right.
\end{eqnarray}

Applying the result in \cite{Digham_2007_TWC}, if the PUs are active, instantaneous SNR is $\upsilon$, and let $Y$ be the output of the integrator in the energy detector, we can derive the conditional pdf, denoted by $f_{Y|\upsilon,\mathcal{H}_{1}}(y)$, for $Y$, which follows the non-central chi-square distribution, i.e.,
\begin{equation}
\label{Idea-H1-PDF}
\begin{aligned}
f_{Y|\upsilon,\mathcal{H}_{1}}(y)=\frac{1}{2}\left(\frac{y}{2\upsilon}\right)^{\frac{m-1}{2}}e^{-\frac{(2\upsilon+y)}{2}}I_{m-1}\left(\sqrt{2\upsilon y}\right),
\end{aligned}
\end{equation}
where $m$ denotes the integer number of samples measured and $I_{\alpha}(\cdot)$ denotes the $\alpha$th-order modified Bessel function of the first kind.

Since the channel is assumed to follow the Rayleigh distribution, the SNR $\upsilon$ follows the exponential distribution with the mean of SNR equal to $\overline{\upsilon}$. Thus, taking into account the fading factor, we have
\begin{equation}
\label{Idea-H1-PDF-Aver-SNR}
\begin{aligned}
f_{Y|\mathcal{H}_{1}}(y)&=\int^{\infty}_{0}f_{Y|\upsilon,\mathcal{H}_{1}}(y)(\overline{\upsilon})^{-1}e^{-\upsilon(\overline{\upsilon})^{-1}}d\upsilon\\
&=\frac{(1+\overline{\upsilon})^{m}e^{-\frac{y}{2(1+\overline{\upsilon})}}}{2(1+\overline{\upsilon})^{2}\overline{\upsilon}^{m-1}}\left[1-\frac{\Gamma(m-1,\frac{\overline{\upsilon}}{2(1+\overline{\upsilon})})}{\Gamma(m-1)}\right],
\end{aligned}
\end{equation}
where $\Gamma(\cdot)$ is the complete gamma function, and $\Gamma(a,z)=\int^{\infty}_{z}t^{a-1}e^{-t}dt$ is the upper incomplete gamma function.

Then, we can derive the cumulative distribution function (CDF), denoted by $F_{Y|\mathcal{H}_{1}}(y)$, for $Y$ given $\mathcal{H}_{1}$ as follows:
\begin{eqnarray}
\label{Idea-H1-CDF}
F_{Y|\mathcal{H}_1}(y)&\!\!=\!\!&\int_0^y f_{Y|\mathcal{H}_1}(t)dt\nonumber \\
&\!\!=\!\!&
\frac{\Gamma(m-1,0)-\Gamma(m-1,\frac{y}{2})}{\Gamma(m-1)}
+\left(\frac{1+\overline{\upsilon}}{\overline{\upsilon}}\right)^{m-1}\nonumber\\
&&\;\;\;\;\;\;\;\cdot
\bigg[1-e^{-\frac{y}{2(1+\overline{\upsilon})}}-
\frac{\Gamma(m-1,0)}{\Gamma(m-1)}\nonumber\\ &&\;\;\;\;\;\;\;
+e^{-\frac{y}{2(1+\overline{\upsilon})}}\frac{\Gamma(m-1,\frac{\overline{\upsilon}y}{2(1+\overline{\upsilon})})}{\Gamma(m-1)}\bigg].
\end{eqnarray}

Therefore, the missed detection probability $p_{MD}$ can be defined and derived as follows:
\begin{equation}
\label{P-MD-Perfect-Full}
\begin{aligned}
p_{MD} \triangleq \mathrm{Pr}\{Y<\beta\;|\;\mathcal{H}_{1}\}=F_{Y|\mathcal{H}_{1}}(\beta),
\end{aligned}
\end{equation}
where $\beta$ is the detection threshold in the perfect full duplex CRNs.

\subsection{The Imperfect Full Duplex CRN}
In the case of imperfect full duplex, $\epsilon^{A}_{ant}$ and $\epsilon^{d}_{ant}$ may not be zero, which is caused by several factors, such as the antenna placement error, the wide bandwidth, the large amplitude difference between the received signal from two transmit antennas, the transmit power, and the channel condition, etc.

The power of self-interference received at receive antenna can be written as follows:
\begin{eqnarray}
\label{Self-Inte-Power}
P_{i}&=&2A_{ant}\left(A_{ant}+\epsilon^{A}_{ant}\right)\left(1-\cos\left(\frac{2 \pi \epsilon^{d}_{ant}}{\lambda}\right)\right)\nonumber \\
&&+\left(d^{A}_{ant}\right)^2.
\end{eqnarray}
It is obvious that if any one or both of $\epsilon^{A}_{ant}$ and $\epsilon^{d}_{ant}$ is not equal to zero, the self-interference power will be not equal to zero. We also map the difference in wavelength to the error in receive antenna placement. Assuming a $q$ difference in wavelength is similar to a $q/4$ error in receive antenna placement. Thereby considering a 20MHz signal centered at 2.48MHz,  $\epsilon^{d}_{ant}$ for 2.47GHz will be $\left[c/\left(2.47\times 10^{6}\right)-c/\left(2.48\times 10^{6}\right)\right]/4$, where $c$ is the speed of light.


When the signal $s(t)$ from the PU is present, using the method in \cite{Urkowitz_1967}, the total received signal of SU, denoted by $r(t)$, can be written as follows:
\begin{eqnarray}
\label{PU-Pres-SU-Sign}
r(t)&=&\sum\limits_{i=1}^{2TW}\bigg[s\left(\frac{i}{2W}\right)+\omega\left(\frac{i}{2W}\right)\nonumber \\
&&+\kappa\left(\frac{i}{2W}\right)\bigg]\mathrm{sinc}\left(2Wt-i\right),
\end{eqnarray}
where $W$ is the bandwidth and $T$ is the observation time interval. Then, we can derive the energy of $r(t)$ in the interval (0, $T$), denoted by $\eta$, as follows:
\begin{eqnarray}
\label{PU-Pres-SU-Sign-Power}
\eta=\int^{T}_{0}r^{2}(t)dt&=&\frac{1}{2W}\sum\limits_{i=1}^{2TW}\bigg[s\left(\frac{i}{2W}\right)\nonumber \\
&&+\omega\left(\frac{i}{2W}\right)+\kappa\left(\frac{i}{2W}\right)\bigg]^{2}.
\end{eqnarray}

We find it convenient to compute the missed detection probability using the output of the integrator over time $T$, denoted by $\widetilde{Y}$, which is a test statistics and can be written as follows:
\begin{equation}
\label{Test-Statistic}
\begin{aligned}
\widetilde{Y} = \frac{\eta}{N_{t}+N_{i}},
\end{aligned}
\end{equation}
where $\eta$ is given by Eq.~\eqref{PU-Pres-SU-Sign-Power}, $N_{t}$ is the two-sided noise power spectral density, $N_{i}$ is the power spectral density of self-interference $\kappa(t)$ and can be written as follows:
\begin{eqnarray}
\label{Self-Inte-PSD}
N_{i}&=&E_{s}-C_{dB}-R_{dB}-D_{dB}-P_{i} \nonumber \\
&=&E_{s}-C_{dB}-R_{dB}-D_{dB}+\left(d^{A}_{ant}\right)^2 \nonumber \\
&&-2A_{ant}\left(A_{amp}+\epsilon^{A}_{ant}\right)\left(1-\cos\left(\frac{2 \pi \epsilon^{d}_{ant}}{\lambda}\right)\right), \nonumber \\
\end{eqnarray}
where $C_{dB}$ is the channel attenuation from the PU to the SU, $R_{dB}$ and $D_{dB}$ are the SNR gains obtained by using the RIC and the DIC techniques, respectively.

We can derive the test statistics $\widetilde{Y}$ in the case when the PU is active can be written as follows:
\begin{equation}
\label{PU-Pres-SU-Sign-Power-Deci}
\begin{aligned}
\widetilde{Y}=\sum\limits_{i=1}^{2TW}\left[\frac{s\left(\frac{i}{2W}\right)+\omega\left(\frac{i}{2W}\right)+\kappa\left(\frac{i}{2W}\right)}{\sqrt{2W\left(N_{t}+N_{i}\right)}}\right]^{2}.
\end{aligned}
\end{equation}

Therefore, the test statistics $\widetilde{Y}$ follows a noncentral chi-square distribution with 2$TW$ degrees of freedom and a non-centrality parameter, denoted by $\delta$, as
\begin{eqnarray}
\label{Non-Para-PU-Pres}
\delta=\sum\limits_{i=1}^{ 2TW}\left[\frac{s\left(\frac{i}{2W}\right)+\kappa\left(\frac{i}{2W}\right)}{\sqrt{2W\left(N_{t}+N_{i}\right)}}\right]^{2}.
\end{eqnarray}
Then, assuming that $s(t)$ and $\kappa(t)$ are uncorrelated over the integration time, Eq.~\eqref{Non-Para-PU-Pres} can be reduced to
\begin{eqnarray}
\label{Non-Para-PU-Pres2}
\delta&=&\sum\limits_{i=1 }^{2TW}\left[\frac{s\left(\frac{i}{2W}\right)}{\sqrt{2W\left(N_{t}+N_{i}\right)}}\right]^{2}+\left[\frac{\kappa\left(\frac{i}{2W}\right)}{\sqrt{2W\left(N_{t}+N_{i}\right)}}\right]^{2}\nonumber\\
&=&\frac{\int^{T}_{0}s^{2}(t)dt}{N_{t}+N_{i}}+\frac{\int^{T}_{0}\kappa^{2}(t)dt}{N_{t}+N_{i}}\nonumber\\
&=&\frac{E_{s}+N_{i}}{N_{t}+N_{i}}.
\end{eqnarray}

We also assume that $\delta$ follows the exponential distribution with the mean equal to $\overline{\delta}$. Then, in imperfect full duplex CRNs, the conditional PDF $\widetilde{f}_{\widetilde{Y}|\frac{\delta}{2},\mathcal{H}_{1}}(y)$ of the test statistic $\widetilde{Y}$ given $\mathcal{H}_{1}$ can be written as follow:
\begin{equation}
\label{Imperfect-H1-PDF}
\begin{aligned}
\widetilde{f}_{\widetilde{Y}|\frac{\delta}{2},\mathcal{H}_{1}}(y)=\frac{1}{2}\left(\frac{y}{\delta}\right)^{\frac{m-1}{2}}e^{-\frac{(\delta+y)}{2}}I_{m-1}\left(\sqrt{\delta y}\right).
\end{aligned}
\end{equation}
Then, we can derive the CDF $\widetilde{F}_{\widetilde{Y}|\mathcal{H}_{1}}(y)$, of $\widetilde{Y}$ given $\mathcal{H}_{1}$ as follows:
\begin{eqnarray}
\label{Imperfect-H1-CDF}
\widetilde{F}_{\widetilde{Y}|\mathcal{H}_{1}}(y)&\!\!=\!\!&\int^{y}_{0}\widetilde{f}_{\widetilde{Y}|\mathcal{H}_{1}}(t)dt\nonumber \\
&\!\!=\!\!&\frac{\Gamma(m-1,0)-\Gamma\left(m-1,\frac{y}{2}\right)}{\Gamma(m-1)}+\left(\frac{1+\overline{k}}{\overline{k}}\right)^{m-1}\nonumber\\
&&\;\;\;\;\;\;\;\cdot\bigg[1-e^{-\frac{y}{2\left(1+\overline{k}\right)}}-\frac{\Gamma(m-1,0)}{\Gamma(m-1)}\nonumber\\
&&\;\;\;\;\;\;\;+e^{-\frac{y}{2(1+\overline{k})}}\frac{\Gamma\left(m-1,\frac{\overline{k}y}{2\left(1+\overline{k}\right)}\right)}{\Gamma(m-1)}\bigg],
\end{eqnarray}
where $\widetilde{f}_{\widetilde{Y}|\mathcal{H}_{1}}(t)=\int^{\infty}_{0}f_{\widetilde{Y}|k,\mathcal{H}_{1}}(t)(\overline{k})^{-1}e^{-k/\overline{k}}dk$, $k=\delta/2$, and $\overline{k}$ is the mean of $\delta/2$.

Therefore, the missed detection probability of imperfect full duplex $\widetilde{p}_{MD}$ can be defined and derived as follows:
\begin{equation}
\label{P-MD-Imperfect-Full}
\begin{aligned}
\widetilde{p}_{MD} \triangleq \mathrm{Pr}\{\widetilde{Y}<\widetilde{\beta}\;|\; \mathcal{H}_{1}\}=\widetilde{F}_{\widetilde{Y}|\mathcal{H}_{1}}(\widetilde{\beta}),
\end{aligned}
\end{equation}
where $\widetilde{\beta}$ is the decision threshold in the imperfect full duplex CRNs. Thus, the packet loss rate of PUs in the imperfect full duplex CRNs can be written as follows:
\begin{equation}
\label{Imperfect-Sensing-Full-Duplex-Frame-Loss-NO}
\begin{aligned}
\rho_{PU} = \widetilde{p}_{MD}.
\end{aligned}
\end{equation}


\section{Simulation Results}
\label{Simulation Results}
\begin{center}
\begin{figure}[!htb]
\centering
\includegraphics[width=3.3 in]{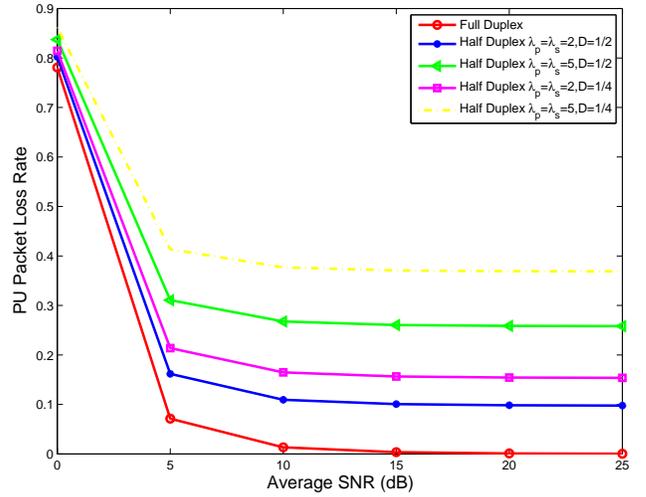}
\caption{The PUs Packet Loss Rate in half duplex CRNs and perfect full duplex CRNs.}
\label{The_PUs_Packet_Loss_Rate}
\end{figure}
\end{center}

To evaluate our developped full duplex CRNs~\cite{Goldsmith_2009}, we simulate the packet loss rate with full duplex CRNs and half duplex CRNs, respectively. Every active PU or SU has $N$ packets to send with $N = 100$ and $L_{p} = 1KByte$. Suppose that $D$ is set to be $N L_{p}/(2R)$ and $N L_{p}/(4R)$, respectively, $\lambda_{p}$ and $\lambda_{s}$ are set to be 2s$^{-1}$ and 5s$^{-1}$, respectively, and $m$ is set to be 5. Fig.~\ref{The_PUs_Packet_Loss_Rate} shows the PUs packet loss rate versus mean SNR in half duplex CRNs and the perfect full duplex CRNs, respectively. From Fig.~\ref{The_PUs_Packet_Loss_Rate}, we can observe that the PU's packet loss rate in full duplex CRNs is lower than that in half duplex CRNs. The PUs' packet loss rate in full duplex CRNs converges to zero when the mean SNR increases. We can also find that the PUs' packet loss also increases when the arrived rate $\lambda_{p}$ and $\lambda_{s}$ increases. The packet loss rates in the cases of $\lambda{p}=\lambda{s}=5, D=1/2$ and $\lambda{p}=\lambda{s}=5, D=1/4$ are higher than our theoretical analyses. This is because in these cases, the channel approaches the full utilization, causing some packets of PUs dropped. Besides, Fig.~\ref{The_PUs_Packet_Loss_Rate} also shows that when the delay bound increases, the PUs' packet loss rate decreases.

\begin{center}
\begin{figure}[!htb]
\centering
\includegraphics[width=3.3 in]{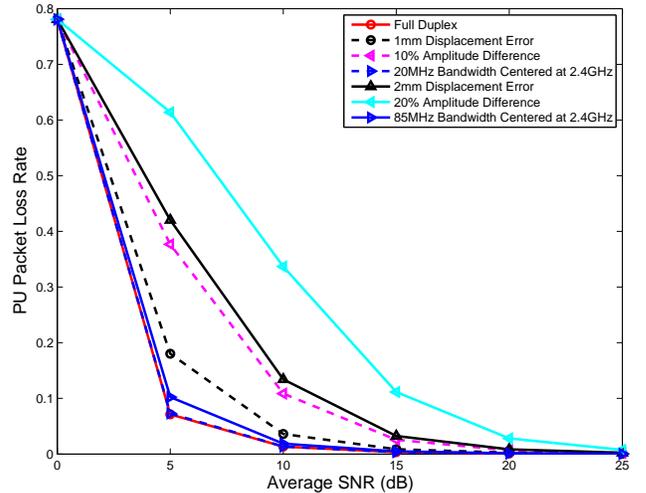}
\caption{The PUs Packet Loss Rate in perfect full duplex CRNs and imperfect full duplex CRNs.}
\label{The_PUs_Packet_Loss_Rate(Imperfect)}
\end{figure}
\end{center}

For imperfect full duplex CRNs, we assume the power of the signal received from the local transmit antenna is -40dBmW, the signal received from the other node is -70dBmW, and the average power of
noise floor is -100dBmW. We consider the effect of receive antenna placement error, amplitude difference between transmit signals from the two transmit antennas, and the bandwidth on full duplex CRNs. Fig.~\ref{The_PUs_Packet_Loss_Rate(Imperfect)} shows that the PUs' packet loss rate versus mean SNR in perfect full duplex CRNs and imperfect full duplex CRNs, respectively. In the imperfect full duplex CRNs, we set $\epsilon^{d}_{ant}$ = 1mm, 2mm, $\epsilon^{A}_{ant} = 0.1A_{ant}, 0.2A_{ant}$, and 20MHz, 85MHz bandwidth centered at 2.48GHz, respectively. From Fig.~\ref{The_PUs_Packet_Loss_Rate(Imperfect)}, we can observe that imperfect full duplex due to the three factors will increase the PUs' packet loss rate. The effect of using wide bandwidth has small effect on the PUs' packet loss rate. The receive placement error and the transmit signals' amplitude difference increase the PUs' packet loss rate slightly. Eliminating the effects of these factors as much as possible will enable the PUs' packet loss rate of the imperfect full duplex CRNs to approach that in perfect full duplex CRNs.

From Fig.~\ref{The_PUs_Packet_Loss_Rate} and Fig.~\ref{The_PUs_Packet_Loss_Rate(Imperfect)}, we can observe that using full duplex in CRNs can effectively decrease the PUs' packet loss rate. But a number of factors will impose the full duplex imperfectness, which causes the PUs' packet loss rate in the full duplex CNRs higher than that in the half duplex CRNs when the mean SNR is low. When the mean SNR is high, even the imperfect full duplex has smaller PUs' packet loss rate than that in wireless half duplex CRNs.

\section{Conclusions}
\label{Conclusions}
We proposed and analyzed the full duplex wireless communication schemes for CNRs by combining the Antenna Cancellation, the RF Interference Cancellation, and the Digital Interference Cancellation techniques for SUs. Compared with using the half duplex CRNs, the SUs can identify the PUs' presence when they are transmitting by using our proposed full duplex CRNs. We used the PUs' packet loss rate to show the advantage of our proposed full duplex over the half duplex in CRNs. We also analyzed the imperfect full duplex in CRNs. Although a number of factors may cause the full duplex imperfect, in high SNR region, the imperfect full duplex has lower PUs' packet loss rate than that in the half duplex CNRs.


\begin{thebibliography}{10}
\bibitem{FCC_2003}
FCC,
\newblock ``Et docket no. 03-237,''
  \url{http://hraunfoss.fcc.gov/edocs\_public/attachmatch/FCC-03-289A1.pdf},
  Nov. 2003.

\bibitem{Hang_CISS_2007}
H.~Su and X.~Zhang,
\newblock ``Opportunistic {MAC} protocols for cognitive radio based wireless
  networks,''
\newblock in {\em IEEE Information Theory Society, the 41st Conference on
  Information Sciences and Systems (CISS 2007)), John Hopkings University},
  Baltimore, MD, USA, Mar 14-16 2007.

\bibitem{Cross_JSAC_XiZhang_2008}
H.~Su and X.~Zhang,
\newblock ``Cross-layer based opportunistic {MAC} protocols for {QoS}
  provisionings over cognitive radio wireless networks,''
\newblock {\em IEEE Journal on Selected Areas in Communications}, vol. 26, no.
  1, pp. 118--129, Jan. 2008.

\bibitem{Peng_2007_WCNC}
P.~Wang, L.~Xiao, S.~Zhou, and J.~Wang,
\newblock ``Optimization of detection time for channel efficiency in cognitive
  radio systems,''
\newblock in {\em Wireless Communications and Networking Conference, 2007.WCNC
  2007. IEEE}, Mar. 2007, pp. 111--115.

\bibitem{Liang_2008_TWC}
Y.~Liang, Y.~Zeng, E.~C.~Y. Peh, and A.~Hoang,
\newblock ``Sensing-throughput tradeoff for cognitive radio networks,''
\newblock {\em IEEE Transactions on Wireless Communications,}, vol. 7, no. 4,
  pp. 1326--1337, Apr. 2008.

\bibitem{Quellan_Inc}
Quellan Inc.,
\newblock ``Qhx220 narrowband noise canceller ic.,''
  \url{http://www.quellan.com/products/qhx220_ic.php.}

\bibitem{Radunovic_WIMESH_2010}
R.~Bozidar, R.~Dinan, K.~Peter, P.~Alexandre, S.~Nikhil, B.~Vlad, and
  D.~Gerald,
\newblock ``Rethinking indoor wireless mesh design: Low power, low frequency,
  full-duplex,''
\newblock in {\em Wireless Mesh Networks (WIMESH 2010), 2010 Fifth IEEE
  Workshop on}, Jun. 2010, pp. 1--6.

\bibitem{Gollakota_Sigcomm_2008}
S.~Gollakota and D.~Katabi,
\newblock ``Zigzag decoding: combating hidden terminals in wireless networks,''
\newblock in {\em Processinds of the ACM SIGCOMM 2008 conference on Data
  communications}, New York, NY, USA, 2008, pp. 159--170.

\bibitem{Choi_2010_Mobicom}
J.~Choi, J.~Mayank, S.~Kannan, L.~Philip, and K~Sachin,
\newblock ``Achieving single channel, full duplex wireless communication,''
\newblock in {\em Proc. 16th ACM MOBICOM}, Chicago, the USA, Sep. 2010.

\bibitem{XiZhang_Multicast_2002}
X.~Zhang, K.~G. Shin, D.~Saha, and D.~Kandlur,
\newblock ``Scalable flow control for multicast abr services in atm networks,''
\newblock {\em IEEE/ACM Transactions on Networking}, vol. 10, no. 1, pp.
  67--85, Feb. 2002.

\bibitem{XiZhang_Multicast_2004}
X.~Zhang and K.~G. Shin,
\newblock ``Markov-chain modeling for multicast signaling delay analysis,''
\newblock {\em IEEE/ACM Transactions on Networking}, vol. 12, no. 4, pp.
  667--680, Aug. 2004.

\bibitem{XiZhang_Flow_2003}
X.~Zhang and K.~G. Shin,
\newblock ``Delay analysis of feedback-synchronization signaling for multicast
  flow control,''
\newblock {\em IEEE/ACM Transactions on Networking}, vol. 11, no. 3, pp.
  436--450, Jun. 2003.

\bibitem{Jia_TWC_2007}
J.~Tang and X.~Zhang,
\newblock ``Cross-layer modeling for quality of service guarantees over
  wireless links,''
\newblock {\em IEEE Transactions on Wireless Communications}, vol. 6, no. 12,
  pp. 4504--4512, Dec. 2007.

\bibitem{Jia_TWC_2_2007}
J.~Tang and X.~Zhang,
\newblock ``Quality-of-service driven power and rate adaptation for
  multichannel communications over wireless links,''
\newblock {\em IEEE Transactions on Wireless Communications}, vol. 6, no. 12,
  pp. 4349--4360, Dec. 2007.

\bibitem{XiZhang_Mag_2006}
X.~Zhang, J.~Tang, H.~H. Chen, S.~Ci, and M.~Guizani,
\newblock ``Cross-layer based modeling for quality of service guarantees in
  mobile wireless networks,''
\newblock {\em IEEE Communications Magazine}, vol. 44, no. 1, pp. 100--106,
  Jan. 2006.

\bibitem{JiaTang_TWC_2007}
J.~Tang and X.~Zhang,
\newblock ``Quality-of-service driven power and rate adaptation over wireless
  links,''
\newblock {\em IEEE Transactions on Wireless Communications}, vol. 6, no. 8,
  pp. 3058--3068, Aug. 2007.

\bibitem{JiaTang_TWC_2008}
J.~Tang and X.~Zhang,
\newblock ``Quality-of-service driven power and rate adaptation over wireless
  links,''
\newblock {\em IEEE Transactions on Wireless Communications}, vol. 7, no. 6,
  pp. 2318--2328, Jun. 2008.

\bibitem{JiaTang_JSAC_2007}
J.~Tang and X.~Zhang,
\newblock ``Cross-layer resource allocation over wireless relay networks for
  quality of service provisioning,''
\newblock {\em IEEE Journal on Selected Areas in Communications (J-SAC)}, vol.
  25, no. 4, pp. 645--656, May 2007.

\bibitem{Cabric_2004_Asilomar}
D.~Cabric, S.~Mishra, and R~Brodersen,
\newblock ``Implementation issues in spectrum sensing for cognitive radios,''
\newblock in {\em Proc. 38th Asilomar Conf. Signals, Syst., Comput.}, Nov.
  2004.

\bibitem{Digham_2007_TWC}
F.~F. Digham, M.~S. Alouini, and M.~K. Simon,
\newblock ``On the energy detection of unknown signals over fading channels,''
\newblock {\em Communications, IEEE Transactions on}, vol. 55, no. 1, pp.
  21--24, Jan. 2007.

\bibitem{Urkowitz_1967}
H.~Urkowitz,
\newblock ``Energy detection of unknown deterministic signals,''
\newblock {\em Proceedings of the IEEE}, vol. 55, no. 4, pp. 523--531, Apr.
  1967.

\bibitem{Goldsmith_2009}
A.~Goldsmith, S.A. Jafar, I.~Maric, and S.~Srinivasa,
\newblock ``Breaking spectrum gridlock with cognitive radios: An information
  theoretic perspective,''
\newblock {\em Proceedings of the IEEE}, vol. 97, no. 5, pp. 894 --914, May
  2009.
\end{thebibliography}

\end{document}